\documentclass[%
reprint,
amsmath,amssymb,
aps,
]{revtex4-2}
\usepackage{graphicx}
\usepackage{dcolumn}
\usepackage{bm}

\begin{document}
	
	\title{Hidden singularities in 3D optical fields}
	\preprint{APS/123-QED}
	
	\author{Xiaoyan Pang}
	\email{xypang@nwpu.edu.cn}
	\affiliation{School of Electronics and Information, Northwestern Polytechnical University, Xi'an, China}%

	\author{Chen Feng}
	\affiliation{School of Electronics and Information, Northwestern Polytechnical University, Xi'an, China}%
	
	\author{Bujinlkham Nyamdorj}
	\affiliation{School of Electronics and Information, Northwestern Polytechnical University, Xi'an, China}%
	
	\author{Xinying Zhao}
	\email{zhaoxinyingxy@163.com.cn}
	\affiliation{School of Physics\&Information Technology, Shaanxi Normal University, Xi'an, China}%
	\affiliation{School of Electronics and Information, Northwestern Polytechnical University, Xi'an, China}%

	\date{\today}



\begin{abstract}
In this article we show that in a three dimensional (3D) optical field there can exist two types of hidden singularities, one is spin density (SD) phase singularity and the other is SD vector singularity, which  are both unique to 3D fields.
The nature of these SD singularities is discussed and their connection with traditional optical singularities is also examined.
Especially it is shown that in a 3D field with purely transverse spin density (`photonic wheels'), these two types of singularities exhibit very interesting behaviors: they are exactly mapped to each other regardless of their different physical meanings and different topological structures.
Our work supplies a fundamental theory for the SD singularities and may provide a new way for further exploration of 3D optical fields.
\end{abstract}
\maketitle
\section{Introduction}
As a new branch of modern optics, singular optics is concerned with optical singularities in the topology of wavefields, which supplies a completely new perspective to study optical fields \cite{Nye,s2001,Dennis}.
Since the seminal papers of the 1970s, the study of singularities has grown significantly over the past decades \cite{gbur2017singular}, and continues to attract a good deal of attention \cite{AP3,AP4,AD10,Pang15,Va16,Ruchi17,Pal18,Pang18p,OAM19,Kotlyar20}. 
The most well-known optical singularities may be phase singularities and polarization singularities, while the coherence singularity is also discussed a lot in a partially coherent field \cite{G2004,G2010p,Pang15} (the coherence singularity sometimes is also treated as a general form of phase singularity).
The study of these singularities has experienced from two dimensional (2D) scalar fields, 2D vector fields to three dimensional (3D) (vector) fields, and in a 3D field some interesting features can be observed, for instance polarization M\"{o}bius strip \cite{Freund2005,Freund2014,Bauer2015}, topological knots~\cite{Dennis2010,Larocque2018}, and Berry's paradox \cite{BP2001,Berry2013,Freund14}.
It is also noted that in a 2D scalar field only the phase singularity can be seen, while in a 2D vector field both the phase singularity and the polarization singularity are observed since an additional freedom in 2D vector field, 
however in a 3D (vector) field, although there is one more freedom, still only these traditional singularities are investigated all the time.
Then a natural question arises: are there any new types of singularity in a 3D vector field?
Before answering this question, let us have a review of the properties of 3D optical fields.

The most distinctive feature of 3D optical fields, like strongly focused fields or near field, is their non-ignorable longitudinal components.
In such a field, the plane of the polarization ellipse is no longer only transverse.
The spin density,  which is short for the spin angular momentum (SAM) density is a quantity to describe the SAM in optical fields,
and also has a very close relationship with polarization state.
In a classical 2D optical field, the spin density is always taken as a scalar because it only has a longitudinal component parallel with the propagation direction.
However, in 3D fields the spin density becomes a vector with its orientation indicating the normal direction of the polarization ellipse \cite{Berry2001,Bliokh2014,neugebauer2015measuring,bekshaev2015transverse,saha2016transverse}.
(From here on, we use `SD' instead of `spin density'.)
It has been found that the SD vector in a 3D field can be spiral along the beam propagation \cite{Pang18,Spin2019,Deng20}, and its
transverse component has a strong connection with geometrical spin Hall effect of light \cite{Aiello2009,neugebauer2014geometric,korger2014observation,Zhu2015,ling2017recent,PRA20}.
Especially, when the spin density vectors are purely transverse, the electric field vectors will spin in the plane containing the propagation direction, so that the `photonic wheels'  occur \cite{banzer2012,Aiello2015,Banzer2016,aiello2016,Chen17,submit}.    
It also has been found that the `photonic wheels'  can have classical topological reactions in the meridional plane of a strongly focused field \cite{submit}.
So actually the spin density vectors manifest the additional freedom of the 3D fields (comparing with 2D vector fields).
Therefore the answer to that question is yes, the hidden singularities in 3D optical fields are the SD singularities.

Here we show that in a 3D optical field, there can exist two new types of optical singularities: the SD phase singularity and the SD vector singularity,  and they are unique to 3D optical fields.
Especially, it will be seen that these two SD singularities can have very interesting properties in the field with `photonic wheels'.

\section{Two types of `hidden' singularities in 3D optical fields}
In a 3D fully polarized field, i.e. the longitudinal polarized component of the field cannot be neglected, the electric field ${\bf E}$ can be written as
\begin{equation}\label{Jone}      
{\bf E} =\left(                
\begin{array}{ccc}   
e_{x}\\e_{y}\\e_{z}  
\end{array}
\right)  =\left(                
\begin{array}{ccc}   
|e_{x}| e^{{\rm i} \phi_x}\\|e_{y}| e^{{\rm i} \phi_y}\\|e_{z}| e^{{\rm i} \phi_z}
\end{array}
\right).                 
\end{equation}
In such a 3D field, 
the polarization ellipses can stay in any plane of the 3D space rather than only in the transverse plane.
Correspondingly, the spin density (SD) is no longer a scalar since the absolute value of the SD of electric field reflects the shape of the polarization ellipse and its direction indicates the orientation of the polarization plane.
The SD vector of electric field, ${\bf s}_E$ in this case is defined as \cite{Bliokh2014,Banzer2016,Berry2001}
\begin{equation}\label{sd}      
{\bf s}_E = \frac{\epsilon_0}{4\omega} {\rm Im} (\mathbf{E}^*\times {\bf{E}})=\left(                
\begin{array}{ccc}   
s^{(x)}_E\\s^{(y)}_E\\s^{(z)}_E  
\end{array}
\right)  =\frac{\epsilon_0}{2\omega}\left(                
\begin{array}{ccc}   
|e_y||e_z|\sin\phi_{zy}\\|e_x||e_z|\sin\phi_{xz}\\|e_x||e_y|\sin\phi_{yx}
\end{array}
\right),     
\end{equation}
with $\epsilon_0$ the permittivity of free space and $\omega$ the angular frequency of the field.
${\rm Im}$ and $^*$ represent the imaginary part and the complex conjugate respectively, and $\phi_{ij}=\phi_i-\phi_j$ ($i,j=x,y,z$) is the phase difference between two field components.
Here we use ${\bf S}_E$ to represent the normalized SD vector and its three components can be expressed as
 \begin{equation}\label{nsd}      
{\bf S}_E =\left(                
\begin{array}{ccc}   
S^{(x)}_E\\S^{(y)}_E\\S^{(z)}_E  
\end{array}
\right)  = \left(                
\begin{array}{ccc}   
s^{(x)}_E/\sqrt{(s^{(x)}_E)^2+(s^{(y)}_E)^2+(s^{(z)}_E)^2}\\s^{(y)}_E/\sqrt{(s^{(x)}_E)^2+(s^{(y)}_E)^2+(s^{(z)}_E)^2}\\s^{(z)}_E/\sqrt{(s^{(x)}_E)^2+(s^{(y)}_E)^2+(s^{(z)}_E)^2}  
\end{array}
\right),    
\end{equation}
thus $|S^{(x)}_E|^2+|S^{(y)}_E|^2+|S^{(z)}_E|^2=1$.
Based on the above theory, in the following we will introduce two types of `hidden' singularities: the SD phase singularity and the SD vector singularity.

\subsection{SD phase singularity}
From Eqs. (\ref{sd}) and (\ref{nsd}), one can find that each component of the SD vector ($S^{(i)}_E$ and $s^{(i)}_E$, $i=x,y,z$) is a real quantity and is also independent of time.
So that by using the method that are adopted for describing the the complex Stokes fields \cite{ST3,ST1,ST2} we can define \textsl{the complex SD fields}, $S^{(ij)}_E$ $(i,j=x,y,z)$ as
\begin{align}
S^{(xy)}_E&=S^{(x)}_E+{\rm i} S^{(y)}_E, \label{sxy}\\
S^{(yz)}_E&=S^{(y)}_E+{\rm i} S^{(z)}_E,\label{syz}\\
S^{(zx)}_E&=S^{(z)}_E+{\rm i} S^{(x)}_E. \label{szx}
\end{align} 
The phase or argument of  $S^{(ij)}_E$, $\psi^{(ij)}_E={\rm arctan}(S^{(i)}_E, S^{(j)}_E)$ denotes the angle of the SD vector in the $i$-$j$ plane from the $+i$-axis with $-\pi<\psi^{(ij)}_E\leq\pi$. 
At the intersection of $S^{(i)}_E=0$ and $S^{(j)}_E=0$, the phase of the complex field $S^{(ij)}_E$ is undefined, 
thus the phase singularity forms and we call this type of singularity the `SD phase singularity'. 

\begin{figure*}[ht]
	\centering
	\includegraphics[width=12.0cm]{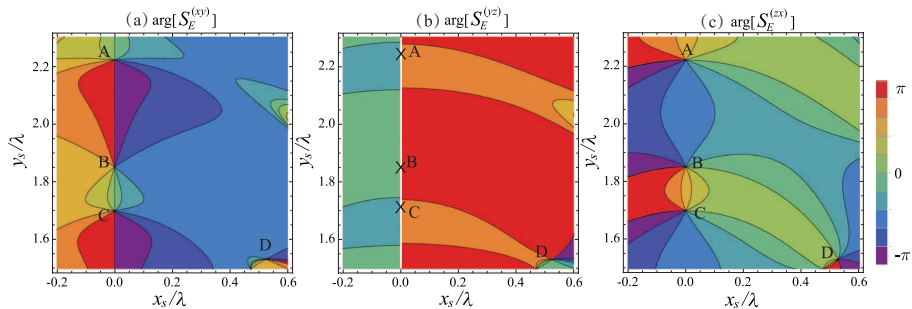}
	\caption{Color-coded plots of the phases of three complex SD fields, (a) $S^{(xy)}_E$, (b) $S^{(yz)}_E$, (c) $S^{(zx)}_E$ at the transverse plane with $z_s=0.5\lambda$ ($\lambda$ is the wavelength). Here the semi-aperture angle of the focusing system is chosen as $\alpha=60^\circ$, and the ratio of the focal length $f$ and the beam waist  $w_0$ is $f/w_0=2$.}
	\label{Fig2}
\end{figure*}

\begin{figure*}[htb]
	\centering
	\includegraphics[width=12.0cm]{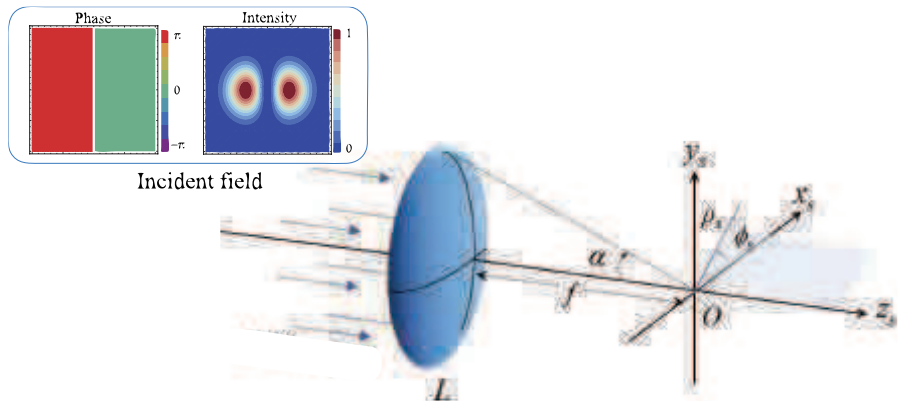}
	\caption{A strongly focusing system with an incident beam that is amplitude tailored.}
	\label{Fig1}
\end{figure*}

The SD phase singularities are illustrated in Fig.~\ref{Fig2}, where the color-coded phases of three complex SD fields, $S^{(xy)}_E$, $S^{(yz)}_E$, $S^{(zx)}_E$ in a general transverse plane of a strongly focused field [see Fig.~\ref{Fig1}, where ($x_s$,$y_s$,$z_s$) are Cartesian coordinates in the focal region] are displayed (the expressions of this 3D optical field will be shown soon).
We can see that in Fig.~\ref{Fig2}(a) and (c), the SD phase singularities of the complex fields $S^{(xy)}_E$ and $S^{(zx)}_E$ are shown at the intersections of different contours, denoted by A, B, C and D. 
It can be observed that the topological charges for these SD singularities are $t=+1$ (at A), $t=-1$ (at B), $t=+1$ (at C) and $t=+1$(at D).
The white line in Fig.~\ref{Fig2}(b) denotes a line of SD phase singularity (i.e. the edge-type or edge singularity) for the complex field $S^{(yz)}_E$, and it can be seen that the phases across the line change a $\pi$.

Here Fig.~\ref{Fig2} is plotted based on the expression of the electric field ${\bf E}(x_s,y_s,z_s)=(e_{x}, e_{y}, e_{z})$ in the focal region of a simple 3D optical model proposed in \cite{submit} (shown in Fig.~\ref{Fig1}), which is written as
\begin{widetext} 
\begin{align}
	e_{x}(\rho_{s},\phi_{s},z_{s})&=\frac{k}{4}\int_{0}^{\alpha}P(\theta)\left[I_{x1}(\theta,\rho_s,\phi_s)+I_{x3}(\theta,\rho_s,\phi_s)\right]e^{\mathrm{i}kz_{s}\cos\theta}\mathrm{d}\theta, \label{Ex}\\
	e_{y}(\rho_{s},\phi_{s},z_{s})&=\frac{k}{4}\int_{0}^{\alpha}P(\theta)\left[I_{y1}(\theta,\rho_s,\phi_s)+I_{y3}(\theta,\rho_s,\phi_s)\right]e^{\mathrm{i}kz_{s}\cos\theta}\mathrm{d}\theta, \\
	e_{z}(\rho_{s},\phi_{s},z_{s})&=-{\rm i}\frac{k}{2}\int_{0}^{\alpha}P(\theta)\left[I_{z0}(\theta,\rho_s,\phi_s)+I_{z2}(\theta,\rho_s,\phi_s)\right]e^{\mathrm{i}kz_{s}\cos\theta}\mathrm{d}\theta,
\end{align}
\end{widetext}
where 
\begin{equation}
	P(\theta)=f^2\sin^2\theta \sqrt{\cos\theta}e^{-f^{2}\sin^{2}\theta/w_{0}^{2}},
\end{equation}
and
\begin{align}
&I_{x1}(\theta,\rho_s,\phi_s)={\rm cos}\phi_{s}(3{\rm cos}\theta+1) J_{1}(k\rho_{s}{\rm sin}\theta),\\
&I_{x3}(\theta,\rho_s,\phi_s)={\rm cos}3\phi_{s}(1-{\rm cos}\theta)J_{3}(k\rho_{s}{\rm sin}\theta), \\
&I_{y1}(\theta,\rho_s,\phi_s)=-{\rm sin}\phi_{s}(1-{\rm cos}\theta) J_{1}(k\rho_{s}{\rm sin}\theta),\\
&I_{y3}(\theta,\rho_s,\phi_s)={\rm sin}3\phi_{s}(1-{\rm cos}\theta) J_{3}(k\rho_{s}{\rm sin}\theta), \\
&I_{z0}(\theta,\rho_s,\phi_s)=-{\rm sin}\theta J_{0}(k\rho_{s}{\rm sin}\theta),\\
&I_{z2}(\theta,\rho_s,\phi_s)={\rm cos}2\phi_{s}{\rm sin}\theta J_{2}(k\rho_{s}{\rm sin}\theta). \label{Iz}
\end{align}
In the above equations $\alpha$, $f$ and $w_0$ are the semi-aperture angle of the focusing system, the focal length and the beam waist of the input beam respectively. 
Also here $k=2\pi/\lambda$ is the wave-number, $J_{n}$ is the first kind of Bessel function of $n$th order and $\rho_{s}=\sqrt{x^{2}_{s}+y^{2}_{s}}$.
Note that in this model the 3D optical field is generated by strongly focusing an $x$-polarized amplitude tailored beam with the complex amplitude distribution  $A_0(\rho,\phi)=\rho{e^{-\rho^2/{w_{0}^{2}}}}{\rm cos}\phi$ \cite{submit} (here $\rho=\sqrt{x^2+y^2}$ and $\phi$ are the radial distance and the azimuthal angle respectively) (see the incident field in Fig. \ref{Fig1}), and  the expressions of $e_{x}$, $e_{y}$ and $e_{z}$ are calculated by the Richards and Wolf vectorial diffraction theory~\cite{RichardWolf}.

From Eqs. (\ref{Ex})-(\ref{Iz}), one can see that except some special regions the three field components $e_{x}$, $e_{y}$ and $e_{z}$ exist in almost all the focal region with different expressions, and their phase differences $\phi_{zy}$, $\phi_{xz}$ and $\phi_{yx}$ vary in 3D space.
This means, according to Eqs. (\ref{sd})-(\ref{szx}),  that the SD phase singularity can be observed easily in a general region of this focused field (for example the transverse plane with $z_s=0.5\lambda$ in Fig.~\ref{Fig2}).
In fact the SD phase singularity is ubiquitous in a space-variant 3D optical field,
because physically a SD phase singularity of the complex field $S^{(ij)}_E$ ($i,j=x,y,z$) means that the spin density vector does not have a component at the $i$-direction or the $j$-direction,
which indicates the polarization ellipse here only can lie in the $i$-$j$ plane.
While it is easy to find such a polarization ellipse in the 3D optical field, i.e. the SD phase singularity is `hidden' in such fields all the time.

Furthermore, Eqs. (\ref{sxy})-(\ref{szx}) show that only two of these complex SD fields are independent, and the third  can be easily got from the others, which also means that  if two kinds of the SD phase singularities, for example the phase singularities of  $S^{(xy)}_E$ and $S^{(yz)}_E$, are coincident at one point,
this point must be a singular point for the third complex field,  $S^{(zx)}_E$.
Physically this is because the only three mutually orthogonal directions in 3D space.
This can be seen clearly in Fig.~\ref{Fig2}, where the points A, B, C and D are the phase singularities of  $S^{(xy)}_E$ and $S^{(yz)}_E$ in Figs.~\ref{Fig2}(a) and (b), so they are also the phase singularities of  $S^{(zx)}_E$ in  Fig.~\ref{Fig2}(c),
while the points except A, B and C on the white line of Fig.~\ref{Fig2}(b) are the phase singularities only for the complex field  $S^{(yz)}_E$.
(Note that the white line in Fig.~\ref{Fig2}(b) denotes that all the points on this line are the SD phase singularities.)

\subsection{SD vector singularity}
In this part we will discuss the second type of `hidden singularity', the `SD vector singularity' or `SD V singularity' in 3D optical fields.

We will first have a very brief review on the traditional vector singularity.
Vector singularity, or V singularity  conventionally describes a point in a electric vector field at which the intensity is null, thus the direction of the electric vector becomes undefined (or all the parameters defining the polarization state are undefined).
Vector singularities are usually discussed in a space-variant linearly polarized field under paraxial regime,
for example in a radially or azimuthally polarized vector field, where a V singularity (i.e. a V point) with Poincar\'{e}-Hopf index, $\eta=+1$ is formed in the beam center.

When an optical field is 3D, the topological structure of the (traditional) vector singularity is not easy to be observed since the electric vector is a time-dependent quantity and the electric field cannot be always linearly polarized in 3D space.
But a 3D optical field also composes a SD vector field,
and in such a vector field the SD vector is time-independent and also `linear' over the whole 3D field except the point at which its SD is null (i.e. $|s^{(x)}_E|=|s^{(y)}_E|=|s^{(z)}_E|=0$).
At the zero SD point, the orientation of the SD vector is undefined (i.e. the `state' of the spin density vector is undefined), thus a vector singularity occurs.
Here we call this vector singularity `SD vector singularity' or `SD V singularity'.
The SD V singularity physically means that the SD (or SAM) is zero there, and at the point with the SD V singularity the polarization of the optical field is linear (or to the extreme the electric field is null).

\begin{figure*}[ht]
	\centering
	\includegraphics[width=12.0cm]{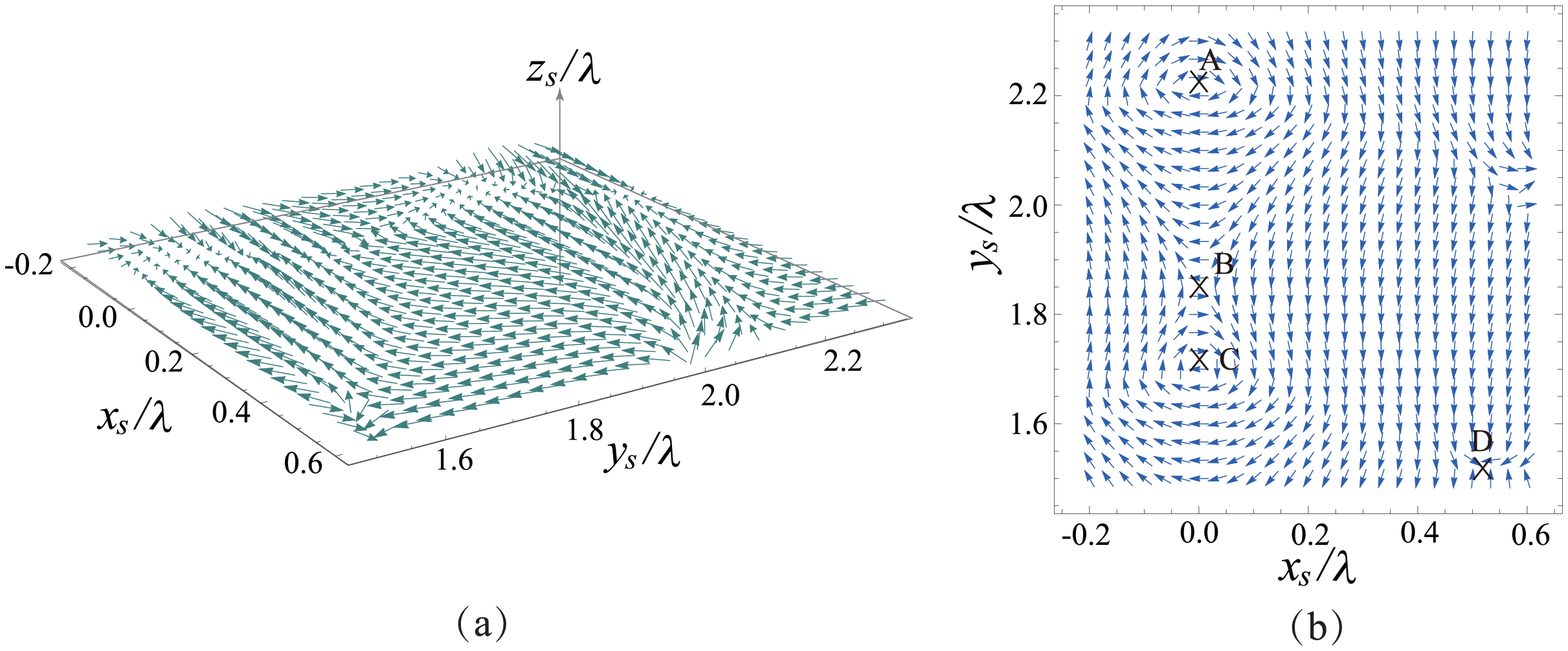}
	\caption{SD vectors of the field with  $z_s=0.5\lambda$. (a) SD vectors in 3D space, (b) Projection of SD vectors on the $x_s$-$y_s$ plane. Here $\alpha=60^\circ$, $f/w_0=2$.}
	\label{Fig3}
\end{figure*}

The SD vectors of the 3D optical field on the same transverse plane as Fig. \ref{Fig2}  are illustrated in Fig.~\ref{Fig3}(a).
Since the topological structures of these 3D vectors are not easy to be observed, their projection on the $x_s$-$y_s$ plane is shown in Fig.~\ref{Fig3}(b).
We can see that at points A, B, C and D the SD V singularities are formed,
and their  Poincar\'{e}-Hopf indices are $\eta=+1$ (at A), $\eta=-1$ (at B), $\eta=+1$ (at C) and $\eta=+1$ (at D).
It is also obvious that the locations of these four SD V singularities are coincident with those of the SD phase singularities in Fig.~\ref{Fig2}. 
The reason for this phenomenon will be discussed in the following.

Let us first make a comparison with the SD phase singularity and the SD V singularity.
The SD phase singularity is formed in the complex SD fields at the point where any two components of SD vector are zero, and this phase singularity can be characterized by the topological charge $t$. 
For a generic topological pattern of a SD phase singularity, the topological charge $t=\pm 1$.
The SD V singularity occurs at a point where all three components of SD vector are null in a SD vector field.
The topological structure surrounding a SD V singularity is usually characterized by the Poincar\'{e}-Hopf index, $\eta$ which is equal to $\pm 1$ for a generic SD V singularity.
Therefore it is quite clear that 
 at the intersection of any two SD phase singularities, for instance $S^{(xy)}_E=0$ and $S^{(yz)}_E=0$,
the SD V singularity is generated, while when a point is a SD V singularity the SD phase singularities must be found there.
In Fig.~\ref{Fig2} the A, B, C and D are the SD phase singularities for both $S^{(xy)}_E$ and $S^{(yz)}_E$ (also for  $S^{(zx)}_E$),
thus at these points the SD V singularities must occur, that is why in Fig.~\ref{Fig3} the SD V singularities are found at these four points. 

In summary, we have introduced two types of `hidden' singularities, the SD phase singularity and the SD V singularity in 3D optical fields.
A strongly focused field is taken as an  example of 3D optical fields, and the basic characteristics of these two types of singularities are investigated in a general plane of this focused field.

In the following part, we will consider some special regions of this focused field and analyze the SD phase singularity and the SD V singularity there.
Especially it will be shown that in certain plane of the field with `photonic wheels' a complex SD field and a SD vector field can become identical, and the SD phase singularities can be exactly mapped to the SD V singularities, although they are very different in physical meaning and topological structure.


\section{SD singularities in the field with `photonic wheels'}
In this section, we will examine the behaviors of these two types of `hidden' singularities in the field of purely transverse spin density (i.e. `photonic wheel').

First let us identify the regions of the field containing `photonic wheels'.
From Eq. (\ref{sd}) we can get that in a field if $s^{(z)}_E=0$ and at least one of the other two SD components, $s^{(x)}_E$ and $s^{(y)}_E$, is not zero,
the purely transverse spin density or `photonic wheel' will be formed in this field.
There are three planes in this focused field [described by Eqs. (\ref{Ex})-(\ref{Iz})] satisfying this condition.
One is the focal plane, and the others are two meridional planes which we will discuss later.
In the focal plane $\phi_{yx}=0,\pi$ [see Eqs.~(\ref{Ex})-(\ref{Iz})], thus from the Eq.~(\ref{sd}), we can get $s^{(z)}_E=0$, whereas $\phi_{zy}, \phi_{xz}=\pm\pi/2$ [see Eqs.~(\ref{Ex})-(\ref{Iz})] and in most regions $|e_x|,|e_y|,|e_z|\neq 0$, so that the `photonic wheels' exist in this plane.

\begin{figure*}[ht]
	\centering
	\includegraphics[width=12.0cm]{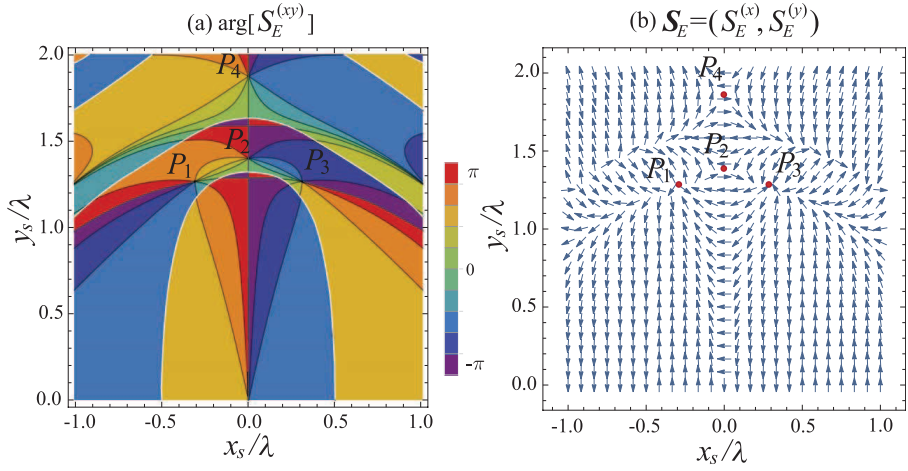}
	\caption{Two types of SD singularities of the field on the focal plane. (a) Color-coded phases of the complex SD fields $S^{(xy)}_E$, (b) SD vectors . Here $\alpha=60^\circ$, $f/w_0=2$.}
	\label{Fig4}
\end{figure*}

In the focal plane, since the `photonic wheels' (i.e. $s^{(z)}_E=0$),
there is only one complex SD field left, i.e., $S^{(xy)}_E=S^{(x)}_E+{\rm i} S^{(y)}_E$,
and the SD vector field degenerates from 3D to 2D, i.e.,  ${\bf S}_E =(S^{(x)}_E,S^{(y)}_E)$.
Thus the complex SD field is now analogous to a complex 2D scalar electric field \cite{F2001},
and the SD vector field becomes a 2D vector field with all the vectors lying in the $x_s$-$y_s$ plane.  
Therefore as it is predicted in \cite{F2001} in this case the complex SD field and the SD vector field can be identical,
and the SD phase singularities can be exactly mapped to the SD vector singularities.
The phase of the complex SD field and the SD vectors are shown in Fig.~\ref{Fig4}.
We can see that in Fig.~\ref{Fig4}(a) there are four (point) SD phase  singularities (i.e. screw dislocations) with their topological charges $t=+1$ (at $P_1$), $t=-1$ (at $P_2$), $t=+1$ (at $P_3$) and $t=-1$ (at $P_4$).
Since these two types of singularities can be mapped to each other, at these four points the SD V singularities also can be observed and their Poincar\'{e}-Hopf indices ($\eta$) equal to the corresponding topological charges ($t$) of the phase singularities at the same points, i.e., $\eta=+1$ (at $P_1$), $\eta=-1$ (at $P_2$), $\eta=+1$ (at $P_3$) and $t=-1$ (at $P_4$) in Fig.~\ref{Fig4}(b) where these four points are marked by red dots.
Note that in Fig.~\ref{Fig4}(a) the white lines denote the line of SD phase singularity, which means that on these lines $S^{(xy)}_E=0$ and the phases have a $\pi$ difference on two sides of these lines,
correspondingly the vectors in Fig.~\ref{Fig4}(b) reverse their directions crossing these lines.

\begin{figure}[ht]
	\centering
	\includegraphics[width=8.0cm]{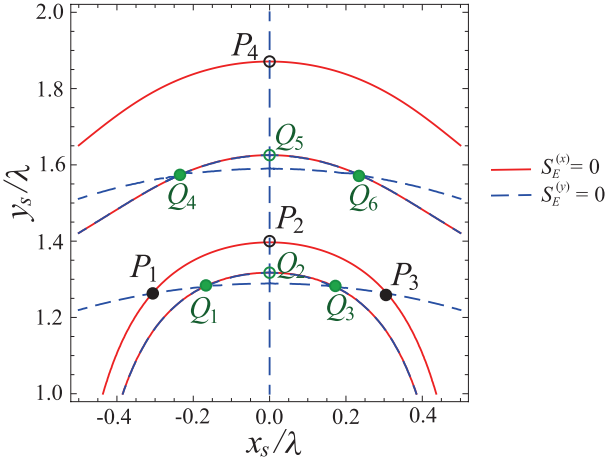}
	\caption{Zero curves of $S^{(x)}_E$ (Re[$S^{(xy)}_E$]) and $S^{(y)}_E=0$ (Im[$S^{(xy)}_E$]) on the focal plane. 
	Positive (negative) charge/index singularities are shown by filled (open) circles. Here $\alpha=60^\circ$, $f/w_0=2$.}
	\label{Fig5}
	\end{figure}

The SD singularities also can be observed through the curves of $S^{(x)}_E=0$ (${\rm Re}[S^{(xy)}_E]=0$) and $S^{(y)}_E=0$ (${\rm Im}[S^{(xy)}_E]=0$), which is shown in Fig.~\ref{Fig5}.
We can see that the SD singularities are located at the points of intersection of the zero crossings $S^{(x)}_E=0$ (solid, red curve) and $S^{(y)}_E=0$ (dashed, blue curve).
Except points $P_1$, $P_2$, $P_3$ and $P_4$ which are typical SD singularities discussed previously,
there also exist other special SD singularities located at the intersection of zero lines.
As one can see, at points  $Q_1$, $Q_2$, $Q_3$, $Q_4$, $Q_5$ and $Q_6$, two zero curves  $S^{(y)}_E=0$ and one zero cure  $S^{(x)}_E=0$ are intersected.
These six singularities are not pure screw dislocations or edge dislocations i.e, they are actually mixed (screw-edge) dislocations (\cite{gbur2017singular}, sec. 3.2).
Through observing these six points in Fig.~\ref{Fig4}, we can get their topological charges/indices as $t=\eta=+1$ (at $Q_1$), $t=\eta=-1$ (at $Q_2$), $t=\eta=+1$ (at $Q_3$), $t=\eta=+1$ (at $Q_4$), $t=\eta=-1$ (at $Q_5$), $t=\eta=+1$ (at $Q_6$).
In Fig.~\ref{Fig5} the filled (open) circles are used to represent the positive (negative) charge/index singularities.

The \textsl{sign rule} says that on every zero crossing of the real or imaginary part of the field, the adjacent singularities  must be of opposite sign ~\cite{F1994,F2001}.
According to this rule, along the zero line ($S^{(y)}_E=0$) $P_4Q_2$, the singularities $Q_5$ and $Q_2$ are `illegal',
also the signs of the singularities at $Q_4$ and $Q_5$ along the curve $Q_4Q_6$ (the blue dashed one) should be different from them  along the curve $Q_4Q_5Q_6$ (the overlapped  curve).
This obvious `violation'  may be explained by the fact that in most researches the sign rule is used to deal with the field with only screw singularities (or the vortices) and all these singularities must obey this sign rule. 
While when a field also has mixed singularities, the topological structure of the whole field are very complicated.
From our observation, in the field with both screw singularity and mixed singularity, these two singularities of different shapes will obey the sign rule in their own `societies', i.e. the screw singularities (or mixed singularities) follow the sign rule in their local regions (these two singularities will not disturb each other). 
This `local sign rule' may be explained as a result of the different distributions of the topological structures of the screw singularity and the edge singularity in 3D space.
In 3D space, (if it is assumed that the transverse plane is chosen as the observation plane) the screw singularity lies in the curve locally parallel to the propagation axis ,
while the edge singularity is located along the curve locally perpendicular to the propagation axis \cite{gbur2017singular} (note that the type of the singularity and its topological structure are dependent on the observer \cite{BP2001,Berry2013,Freund14}).
However, the topological charge of a phase singularity is defined in a 2D plane (i.e. the observation plane),
thus the singularities along two mutually orthogonal curves hardly react with each other topologically.
So the mixed singularities and the screw singularities in present case are also not easy to `disturb' each other.
In a certain sense, this `local sign rule' is a manifestation of the Berry's paradox in 3D fields \cite{BP2001,Berry2013,Freund14}. 
According to this `local sign rule',
in Fig.~\ref{Fig5}, we can see that in the region without mixed singularity, the screw singularities at points $P_1$, $P_2$ and $P_3$ alter their signs along the zero curve of $S^{(x)}_E$ (the solid, red one), while in the region without screw singularities, the mixed singularities at $Q_1$, $Q_2$ and $Q_3$ (or $Q_4$, $Q_5$ and $Q_6$) also obey the sign rule. 
Actually the sign rule is usually used to guarantee the topological conservation for phase/vector singularities in their topological behaviors.
In the following we will show that the topological reactions for these two SD phase singularities of different shapes (and their corresponding SD V singularities) can be observed in the focal plane, and during their reaction process all the singularities obey the `local sign rule' and their topological charges/indices are conserved. 

\begin{figure*}[htb]
	\centering
	\includegraphics[width=13cm]{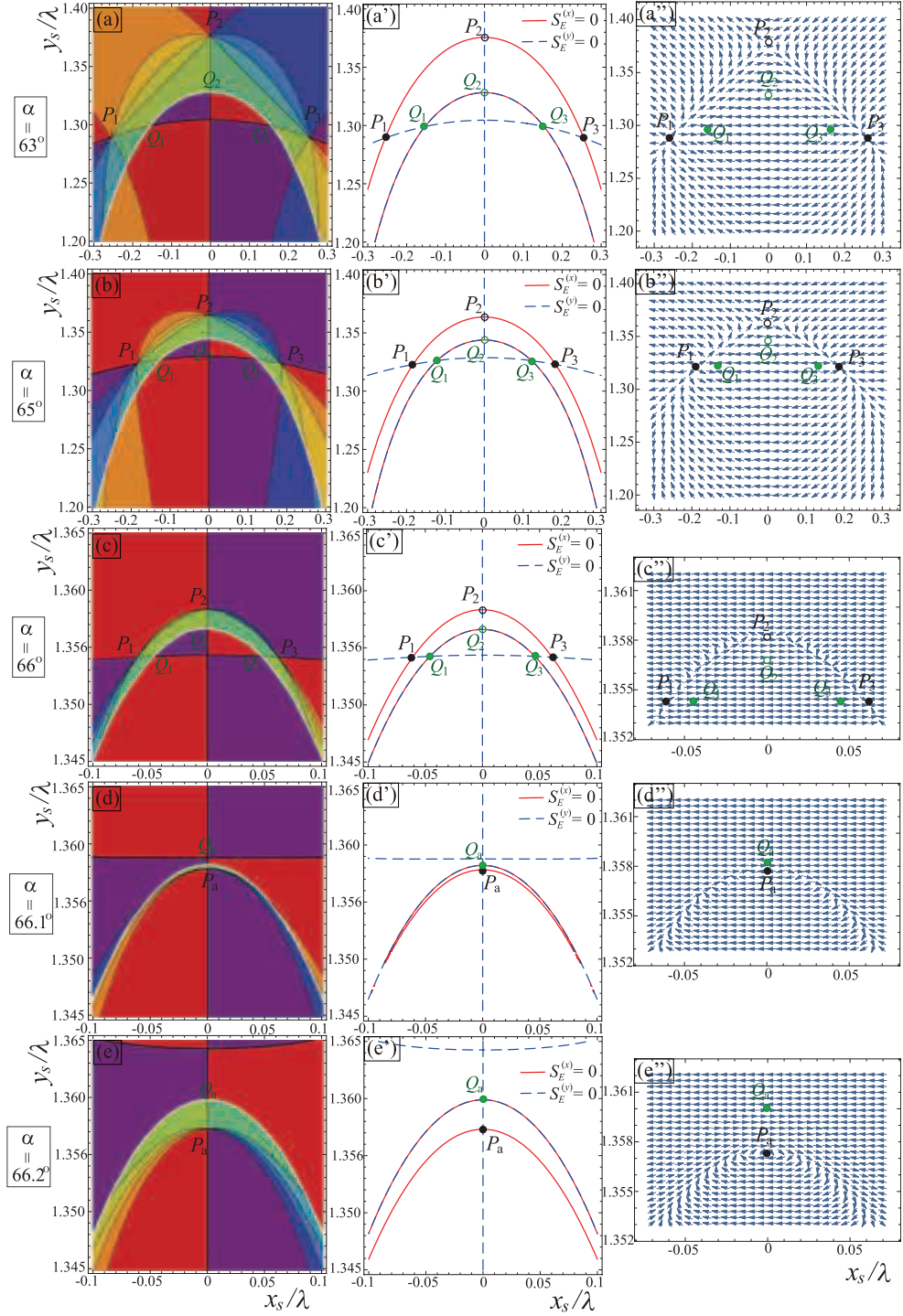}
	\caption{Topological reaction of the SD phase/vector singularities in the focal plane. Here $f/w_0=2$.}
	\label{Fig6}
\end{figure*}

In order to observe 
the topological behaviors more clearly, here we only focus on the singularities at points $P_1$, $P_2$, $P_3$ and $Q_1$, $Q_2$, $Q_3$.
The topological reaction, the annihilation/creation event of SD singularities is presented in Fig.~\ref{Fig6},
where in the left column the phase variation of the complex SD field $S^{(xy)}_E$ is shown, in the middle column the events are expressed by the behaviors of zero curves of Re[$S^{(xy)}_E$] and Im[$S^{(xy)}_E$], and the right column displays the corresponding changes of SD vectors in the same region.
In the plots of the first line, the semi-aperture angle $\alpha$ is set as $63^\circ$,
and in the plots of following lines $\alpha$ is increased slowly from $65^\circ$, $66^\circ$, $66.1^\circ$ to $66.2^\circ$.
In all plots of Fig.~\ref{Fig6} we have $f/w_0=2$.
With the increase of $\alpha$, we can see first that points $P_1$, $P_2$ and $P_3$ become more and more closer to each other, and this is the same for points $Q_1$, $Q_2$ and $Q_3$ (note that because when $\alpha=66^\circ\sim 66.2^\circ$ these singularities are confined in a very small region, so the observation plane is also narrowed in the plots of lines 3, 4 and 5); 
second the zero curve $P_1P_2P_3$ moves down while the zero curves $Q_1Q_2Q_3$ moves up, thus these six SD singularities come closer and closer from plot (b) [(b')(b'')], (c)[(c')(c'')] to (d) [(d')(d'')].
When $\alpha=66.1^\circ$, three SD singularities at points $P_1$ with $t=+1$, $P_2$ with $t=-1$ and $P_3$ with $t=+1$ (at points $Q_1$ with $t=+1$, $Q_2$ with $t=-1$ and $Q_3$ with $t=+1$) annihilate partially at point $P_a$ ($Q_a$).
As we can see in plot (d), a new SD screw singularity (a new SD mixed singularity) with topological charge $t=+1$ at point $P_a$ ($t=+1$ at point $Q_a$) is formed.
For the SD vectors, since it is not easy to get enough vectors around the singularities involved in the reaction,
this annihilation event could not be observed clearly in plot (d''). 
As $\alpha$ increases slightly from $66.1^\circ$ to $66.2^\circ$, the zero curve with point $P_a$ still moves down whereas the zero curve with $Q_a$ moves up, so that the two new born singularities are separated by a distance which is more obvious to be seen [see plot (e) (e') and (e'')].
If $\alpha$ decreases from $66.2^\circ$ to $63^\circ$, this annihilation process will be `played' in reverse, i.e. the creation of SD singularities can be observed.
During this annihilation/creation process, we can see that the topological charge/index is conserved  and all the SD singularities obey the `local sign rule' i.e., the screw singularities (or mixed singularities) interact with each other in their own society.
It is also confirmed again that in the field with `photonic wheels' the two types of SD singularities can be exactly mapped to each other. The SD complex field $S^{(xy)}_E$ and the SD vectors in the focal plane experience the same topological reactions but through different forms.

\section{Connections with traditional optical singularities}
In this section, the relation between the SD singularities  and traditional singularities of electric fields will be investigated.

\begin{figure*}[htb]
	\centering
	\includegraphics[width=12.0cm]{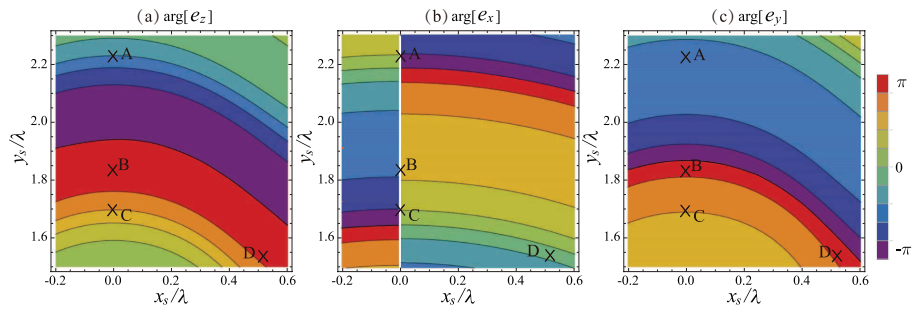}
	\caption{Color-coded plots of the phases of three electric field components, (a) $e_z$, (b) $e_x$, (c) $e_y$ at the transverse plane $z_s=0.5\lambda$. Here $\alpha=60^\circ$, $f/w_0=2$.}
	\label{Fig7}
\end{figure*}

Firstly, we will show that the traditional phase singularities can induce the SD phase singularities.
From Eqs. (\ref{sd}) and (\ref{sxy}) we can get 
\begin{equation}\label{sme}
S^{(xy)}_E=S^{(x)}_E+{\rm i} S^{(y)}_E=M_e \left(|e_y|\sin\phi_{zy}+{\rm i}|e_x|\sin\phi_{xz} \right)|e_z|
\end{equation}
here $M_e=\epsilon_0/2\omega\sqrt{(s^{(x)}_E)^2+(s^{(y)}_E)^2+(s^{(z)}_E)^2}$. 
This means that if at a point $|e_z|=0$, $S^{(x)}_E=S^{(y)}_E=0$ and a SD phase singularity of the complex field $S^{(xy)}_E$ will occur at this point. 
Similarly, we can get that the point of $|e_x|=0$ ($|e_y|=0$) must be a SD singularity of the complex field $S^{(yz)}_E$ ($S^{(zx)}_E$).
While $|e_i|=0$ ($i=x,y,z$) also means the phase singularity of the electric field component $e_i$.
Therefore, we can say that when a point is a phase singularity of the electric field $e_i$, it must be a SD phase singularity of the complex field $S^{(jk)}_E$ ($j,k=x,y,z$ and $j\neq k \neq i$).
However, Eq. (\ref{sme}) also implies that the SD phase singularity can happen without the condition $|e_z|= 0$ (for instance $\sin\phi_{zy}=\sin\phi_{xz}=0$ can induce the SD phase singularity).
Then a SD phase singularity observed at a point does not mean that the traditional phase singularity of the electric field can also be seen there.
Let us see one example.
Fig.~\ref{Fig2} shows the phases of three complex SD fields, and points A, B, C and D in this figure represent the four SD phase singularities. 
The white line in Fig.~\ref{Fig2}(b) denotes the line of SD phase singularity of the field $S^{(yz)}_E$.
The phases of the corresponding electric field components $e_x$, $e_y$ and $e_z$ of these three complex SD fields are illustrated in Fig.~\ref{Fig7}, where the white line denotes the line of phase singularity of $e_x$.
First, we can see the location of the white line in Fig.~\ref{Fig7}(b) is coincident with that of the white line in Fig.~\ref{Fig2}(b), which shows that the phase singularities of $e_x$ are also the SD phase singularities of $S^{(yz)}_E$.
Second, the SD phase singularities at points A, B, C and D are not phase singularities of the electric field components $e_y$ and $e_z$, especially at the point D no phase singularity is formed for any electric field component.   

Secondly, the SD V singularity has a close relation with the traditional polarization singularity.
There are two types of polarization singularity: L-type and C-type.
A L-type singularity of a polarization variant field occurs at the point with linear polarization (i.e. the handedness of the polarization ellipse is undefined);
while a C-type singularity is formed at the point where the field is circularly polarized (i.e. the orientation of the polarization ellipse is undefined) \cite{AD02,AD022,AD10,gbur2017singular}.
From Eq.~\ref{sd}, one can see that if at a point $|{\bf s}_E|=0$, this point must be a L-point (linearly polarized), 
while if  $|{\bf s}_E|$ gets its maximum, it should be a C-point.
Thus when a point is a SD V singularity, it must be a L-type polarization singularity, and vice versa. 
For example, the white curves in Fig.~\ref{Fig4}(a) are actually the L-lines of the electric field with $|e_z|=0$ (the fields there are linearly polarized at $\rho_s$-direction), 
and $P_1$, $P_2$, $P_3$ and $P_4$ are also L-points with the electric fields polarized at $z_s$ direction ($|e_\rho|=0$).
Note that at the points of the traditional V singularity the intensity of the electric field is zero, which also means at these points the SD V singularity are also formed.

In addition, based on the relations discussed above one can see some special regions existing both the SD singularities and traditional singularities in this focused field .
 One is the propagation axis, along which only $e_z$ component is nonzero [Eqs. (\ref{Ex})-(\ref{Iz})], so that $|{\bf s}_E|=0$, thus the $z_s$ axis is the L-line of the electric field, and is also a line of SD V singularity.
The two meridional planes, the $x_s$-$z_s$ plane and the $y_s$-$z_s$ plane, as we mentioned before, are also special because the `photonic wheels' there.
We will first briefly demonstrate that in these two meridional planes the `photonic wheels' exist.
 On the $x_s$-$z_s$ plane, $\phi_s=0,\pi$, thus from  Eqs. (\ref{Ex})-(\ref{Iz}) we can get that  $e_y=0$.
Then since Eq. (\ref{sd}),  $s^{(z)}_E=0$ (also $s^{(x)}_E=0$), which means that spin density vectors in this plane are purely transverse, i.e. `photonic wheels' are formed.
Similarly, it also can be got that on the $y_s$-$z_s$ plane the `photonic wheels' exist.
Because $e_y=0$ and $s^{(x)}_E=s^{(z)}_E=0$ on the $x_s$-$z_s$ plane, 
this meridional plane is both a surface of phase singularity of the electric field $e_y$ and a surface of the SD phase singularity of the complex field $S^{(zx)}_E$.
By the same way, we can get that the $y_s$-$z_s$ plane is also both a surface of phase singularity of the electric field component $e_x$ and a surface of  SD phase singularity of the complex field $S^{(yz)}_E$.

\section{Conclusions}
In this article, we propose two new types of singularities which can exist in a 3D optical field, the SD phase singularity and the SD V singularity.
The topological properties of these SD singularities and their connection with traditional optical singularities are discussed,
also the point/line/surface of the SD singularities are observed in a strongly focused field.
It is found that the SD singularities obey the `local sign rule' during their topological reactions, i.e. the topological charges/indices are conserved.
It is also shown that the traditional phase singularity can induce the SD phase singularity, but not the way around, while the SD V singularity and the L-type polarization singularity always occur at the same place.
In a field of purely transverse SD (i.e. `photonic wheels'), we find that there exists  a one-to-one mapping between the SD phase singularity and the SD V singularity regardless of their different manifestations, which confirms the theory of the relation between the phase singularity and vector singularity predicted in \cite{F2001}.
Our research will add new types of optical singularities in singular optics, and may provide another perspective to explore the special properties of 3D optical fields.
Furthermore, the rotational topological structures around the SD singularities and their topological behaviors discussed in this article 
can be applied in the spin-dependent optical manipulations \cite{AP1} and in the optical tweezers \cite{AP2}. 
The special behaviors of the SD singularities in the field with `photonic wheels' will be useful in the application of the transverse SD, for instance controlling light-matter interactions on the level of individual atoms \cite{banzer2012,Aiello2015}.
Our findings may also have implications for the observation of traditional optical singularities in a 3D field \cite{AP3,AP4,Dennis2010,Larocque2018}.

\section*{Funding}
National Natural Science Foundation of
China (NSFC) (No. 11974281,11504296), Natural Science Basic Research Plan in Shaanxi Province of China (No. 2020JM-116).

\section*{Disclosures}
The authors declare that there are no conflicts of interest related to this article. 


\bibliography{Hid}

\end{document}